# Physics-informed neural network surrogate modeling of single particle model for lithium-ion batteries

Yi Zhuang*, Yusheng Zheng*, Yunhong Che*, Remus Teodorescu*

*Aalborg University, Aalborg, 9220, Denmark (yizhu@energy.aau.dk, yzhe@energy.aau.dk, ych@energy.aau.dk, ret@energy.aau.dk )

**Abstract**: Physics-based models play a key role in battery management, yet face challenges in real-time applications due to the high computational cost of solving coupled algebraic-partial differential equations. To accelerate model simulation, this study benchmarks three physics-informed neural network (PINN) architectures for modeling the battery single particle model, including two conventional PINN architectures and a DeepONet-based architecture. Both the accuracy and the generalization of these PINNs have been evaluated and compared under various current conditions. Our results highlight the potential of PINNs in modeling battery physics but also reveal limitations of conventional PINN architectures under highly dynamic current conditions. Among them, the Fourier-enhanced DeepONet achieves superior generalization performance and offers nearly a 10× speedup compared with numerical solvers. This work provides an example of integrating physics-based models with data-driven solutions to not only accelerate computation but also provide more physical insights.

## 1. INTRODUCTION

Lithium-ion battery pack serves as the primary power and energy storage component for electric vehicles, where hundreds or even thousands of cells are assembled. As a result, the battery management system (BMS) plays a critical role in ensuring operational safety, efficiency, and reliability (Li et al., 2025). Traditional BMS implementations rely on equivalent circuit models (ECMs) for state estimation or optimal control. Although these empirical models are computationally efficient, they struggle to capture the complex electrochemical processes inside the cells during operation. This limitation fundamentally restricts the long-term accuracy and robustness of ECM-based estimation methods. In contrast, physics-based models (PBMs) offer a promising path toward next-generation BMS technologies. Based on first-principles electrochemical theory, PBMs describe internal battery states through coupled partial differential equations and algebraic equations, enabling high-fidelity prediction of concentration dynamics, reaction kinetics, and internal potentials. However, the strong nonlinearity and multi-scale coupling inherent in these models pose significant challenges for real-time control and onboard implementation, necessitating various model-reduction strategies. As summarized in Li et al. (2022), existing reduction approaches are primarily based on mathematical simplifications.

In recent years, advances in machine learning have opened new opportunities for reconstructing physics-based models with significantly improved computational efficiency. Compared with traditional reduced-order modeling techniques, data-driven and physics-informed learning methods offer the potential to preserve essential electrochemical behavior while enabling fast, real-time computation, thus providing a powerful alternative for next-generation BMS applications.

Physics-informed learning methods (Raissi et al., 2019) have gained increasing attention due to their data efficiency and strong generalization capability. Hassanaly et al. (2024a, 2024b) developed physics-informed neural network (PINN) based surrogate models for both the single particle model (SPM) and pseudo-2D (P2D) models without using any labeled data, demonstrating the strong potential of PINNs for P2D modeling. Lee et al. (2025) also constructed a P2D surrogate model using PINNs and proposed a method to address the nonlinearity of the Butler–Volmer equation. However, a major limitation of these models is that they take only time and space as inputs, which restricts their applicability to the specific current profiles used during training. To further improve the generalization ability of PINNs, operator-learning frameworks such as PI-DeepONet (Lu et al., 2021a; Wang et al., 2021a) and PI-FNO (Li et al., 2024) were introduced. These methods learn mappings between function spaces and therefore exhibit much stronger generalization performance. Zheng et al. (2023) applied PI-DeepONet to construct an enhanced single particle model (ESPM) surrogate model, while Brendel et al. (2025) combined PI-DeepONet with Physics-Parameterized PINNs (Cho et al., 2024) and spatio-temporal multi-scale Fourier features (Wang et al., 2021b) to further enhance generalization.

Despite these advances, existing studies still lack a systematic investigation into the generalization limitations of conventional PINN architectures, as well as how labeled data affect PINN performance. Moreover, direct comparisons between conventional PINNs and operator-learning approaches such as PI-DeepONet remain sparse, leaving open questions about their relative strengths and applicability to battery modeling.

To address these gaps, this work makes two primary contributions. First, we designed new PINN architectures that include the current profile as an input, thereby improving its generalizability at varying conditions. To demonstrate the data efficiency, we also investigate how different amounts of labeled data influence model accuracy. Second, we provide a comprehensive comparison between three PINN architectures in terms of their performance and generalization capabilities, thus guiding the design of PINN architecture for modeling more complex battery physics.

## 2. PHYSICS-BASED MODEL

### 2.1 Single particle model

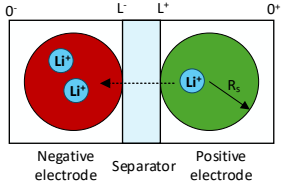

Figure 1. Schematic of the SPM

Different from the P2D model, which includes multiple particles distributed along the x-scale, the SPM assumes a uniform molar flux density $j_n$ across the entire x-scale. Therefore, each electrode is represented by a single particle, and $j_n$ can be calculated directly as (Zhang et al., 2000):

$$j_n^+(t) = -\frac{I(t)}{Aa_s^+ FL^+}, \quad j_n^-(t) = \frac{I(t)}{Aa_s^- FL^-} \quad (1)$$

where $I(t)$ is input current, $a_s$ is the specific interfacial surface area $a_s = 3\varepsilon_s/R_s$.

The diffusion of lithium-ion concentration within the particle is governed by:

$$\frac{\partial c_s(r,t)}{\partial t} = \frac{1}{r^2}\frac{\partial}{\partial r}\left(r^2 D_s \frac{\partial c_s(r,t)}{\partial r}\right) \quad (2)$$

where $r$ denotes the radial coordinate of the active-material particles. The boundary and initial conditions for (2) are given by:

$$c_s(r,0) = c_s^0$$

$$\left.\frac{\partial c_s}{\partial r}\right|_{r=0} = 0$$

$$\left.D_s \frac{\partial c_s}{\partial r}\right|_{r=R_s} = -j_n(t)$$

The overpotential is calculated using the Butler-Volmer equation:

$$j_n(t) = \frac{2i_0}{F}\sinh\left(\frac{F}{2RT}\eta(t)\right) \quad (3)$$

where $\eta(t)$ is the reaction overpotential and $i_0(t)$ is the exchange current density, given by:

$$i_0(t) = k\bar{c}_e^{0.5} c_{ss}^{0.5}(t)\left(c_{s,max} - c_{ss}(t)\right)^{0.5}$$

Then, the terminal voltage can be expressed as the solid-phase electric potential difference between the positive and negative electrodes:

$$V(t) = \Phi_s(0^+, t) - \Phi_s(0^-, t) \quad (4)$$

The electric potential in solid electrode is given by:

$$\Phi_s(t) = U_{OCP}\left(c_{ss}(t)/c_{s,max}\right) - \eta(t) \quad (5)$$

where $c_{ss}(t)$ is the particle surface concentration.

The meanings and corresponding values of these model parameters are presented in Table A1.

## 3. SURROGATE MODEL FOR SPM

Physics-based battery models typically involve partial differential equations (PDEs), which significantly slow down numerical simulation. To enable real-time implementation, we employ the concept of PINN to solve the PDEs embedded in the SPM, resulting in a surrogate model referred to as the physics-informed SPM (PI-SPM). In this section, we provide a concise introduction to PINNs, describe the network architectures that can be employed, and discuss how nondimensionalization and hard constraints can be incorporated to facilitate more efficient training.

### 3.1 Physics-informed neural network

PINNs constitute a class of deep learning models that embed the governing physical laws, typically expressed as PDEs, directly into the training objective. Consider a general time-dependent PDE defined in a spatiotemporal domain $\Omega \subset \mathbb{R}^d \times [0, T]$:

$$\mathcal{N}[u(x,t)] = 0, (x,t) \in \Omega,$$

subject to the initial and boundary conditions

$$u(x,0) = h(x), \quad u(x,t) = g(x,t), \quad (x,t) \in \partial\Omega.$$

Here, $x$ and $t$ denote the spatial and temporal coordinates, $\mathcal{N}$ is a general linear or nonlinear differential operator, $u(x,t)$ represents the PDE solution, $h(x)$ is the prescribed initial condition, and $g(x,t)$ defines the boundary condition on the boundary $\partial\Omega$. This formulation is general and can be readily extended to higher-order PDEs by rewriting them as systems of first-order equations.

To approximate the unknown solution $u(x,t)$, a fully connected neural network $u_{NN}(x,t)$ is constructed. Automatic differentiation—now widely available in modern machine-learning frameworks—enables exact, mesh-free evaluation of the differential operator $\mathcal{N}[u_{NN}]$, thus avoiding discretization errors and greatly simplifying implementation. During training, the network parameters are optimized to minimize a composite loss function that enforces the physical constraints encoded by the PDE, while also incorporating initial conditions, boundary conditions, and optionally observational data. The resulting loss function can be written as:

$$\mathcal{L} = \lambda_r \mathcal{L}_r + \lambda_b \mathcal{L}_b + \lambda_0 \mathcal{L}_0 + \lambda_d \mathcal{L}_d,$$

where

$$\mathcal{L}_r = \frac{1}{N_r}\sum_{i=1}^{N_r} \| \mathcal{N}[u_{NN}(x_i,t_i)] \|^2,$$

$$\mathcal{L}_b = \frac{1}{N_b}\sum_{i=1}^{N_b} \| u_{NN}(x_i,t_i) - g(x_i,t_i) \|^2,$$

$$\mathcal{L}_0 = \frac{1}{N_0}\sum_{i=1}^{N_0} \| u_{NN}(x_i,0) - h(x_i) \|^2,$$

$$\mathcal{L}_d = \frac{1}{N_d}\sum_{i=1}^{N_d} \| u_{NN}(x_i,t_i) - u_i^{data} \|^2.$$

In these expressions, $N_r$, $N_b$, $N_0$, and $N_d$ denote the numbers of residual, boundary, initial, and data points, respectively, whereas $\lambda_r, \lambda_b, \lambda_0, \lambda_d$ are weighting coefficients balancing the different loss components. Minimizing $\mathcal{L}$ ensures that the learned solution satisfies both observational data and governing physics. PINNs have demonstrated strong potential in solving forward and inverse problems, parameter estimation, and uncertainty quantification. Their ability to unify data-driven learning with physical constraints makes them a powerful tool for scientific machine learning and computational modeling.

### 3.2 Model architecture

As illustrated in Fig. 2, three different model architectures are constructed: a split architecture, a merged architecture, and a DeepONet-based architecture. The first two architectures are based on conventional PINN structures, while the third corresponds to the PI-DeepONet model. All three architectures have the same inputs: time, space, and current. In the split architecture, the two concentration states are predicted independently using two separate neural networks. The advantage of this configuration is that the network parameters are not shared, allowing each network to better capture the specific characteristics of its corresponding state. In contrast, the merged architecture employs a single neural network to predict both concentrations simultaneously. Since the two states satisfy similar governing equations, sharing parameters enables a more compact architecture and reduces computational complexity. The third architecture is based on DeepONet, where the modification is introduced by adding a linear layer (the blue block in Fig. 2) after combining the outputs of the branch and trunk networks, enabling the prediction of both states. Because DeepONet is designed to learn mappings between input and output functions (i.e., operators), it is expected to provide superior performance when handling dynamic current inputs.

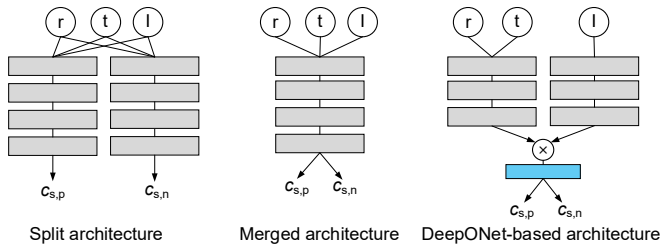

Figure 2. Architecture of the PI-SPM.

### 3.3 Nondimensionalization

Data normalization is an important preprocessing step in machine learning, which involves transferring the input features and output into a similar scale, and can make the training process more stable and converge faster. In the SPM, there is one microscopic coordinate ($r$) and a temporal coordinate ($t$). These multi-scale features often make the PINN training fail to converge (Wang et al., 2022). Similar to the data/features normalization, the nondimensionalization technique in PINN has been proven to have a significant impact on model accuracy and convergence speed (Wang et al., 2023). In the nondimensionalization process of SPM, the coordinates $t$, $r$, and concentration $c$ are rescaled according to characteristic reference values as follows:

$$\tilde{t} = \frac{t}{(R_s^-)^2/D_s^-}, \qquad \tilde{r} = \frac{r}{R_s}, \qquad \tilde{c}_s = \frac{c_s}{c_{s,max}}$$

Following this nondimensionalization, the rescaled governing equation (2) and its corresponding initial and boundary conditions are given by:

$$\frac{\partial \tilde{c}_s^+}{\partial \tilde{t}}\tilde{r} = \left(\frac{\partial \tilde{c}_s^{+2}}{\partial^2 \tilde{r}}\tilde{r} + 2\frac{\partial \tilde{c}_s^+}{\partial \tilde{r}}\right)\cdot \frac{D_s^+}{D_s^-}\frac{(R_s^-)^2}{(R_s^+)^2}$$

$$\frac{\partial \tilde{c}_s^-}{\partial \tilde{t}}\tilde{r} = \left(\frac{\partial \tilde{c}_s^{-2}}{\partial^2 \tilde{r}}\tilde{r} + 2\frac{\partial \tilde{c}_s^-}{\partial \tilde{r}}\right)$$

$$\left.\frac{\partial \tilde{c}_s}{\partial \tilde{r}}\right|_{\tilde{r}=0} = 0$$

$$\left.\frac{\partial \tilde{c}_s}{\partial \tilde{r}}\right|_{\tilde{r}=1} = -j_n \frac{R_s}{D_s c_{s,max}}$$

To avoid the numerical instability introduced by the $1/r$ term at $r = 0$, both sides of the equation are multiplied by $r$.

### 3.4 Initial condition hard constraint

In the vanilla PINN framework, the initial and boundary conditions are imposed by minimizing their residuals within the loss function, a strategy commonly referred to as the *soft-constraint* method. Although widely used and often effective, incorporating these residual terms inevitably increases the training complexity. To alleviate this issue, several studies (Lu et al., 2021b; Sukumar and Srivastava, 2022) have proposed *hard-constraint* methods, in which the neural network outputs are reconstructed such that the initial and boundary conditions are inherently satisfied.

In our work, the output of the neural network $NN(\tilde{r}, \tilde{t}, I; \theta)$ is passed through an output-transformation layer to ensure that the initial condition $\tilde{c}(\tilde{r}, 0) = \tilde{c}_0$ is exactly satisfied. Specifically, the transformed output is defined as

$$\tilde{c}(\tilde{r}, \tilde{t}) = NN(\tilde{r}, \tilde{t}, I; \theta)\,\tilde{t} + \tilde{c}_0,$$

which enforces the initial condition by construction.

## 4. RESULTS

In this section, we present a comprehensive comparison of the three surrogate models under both constant-current and dynamic-current conditions. The reference model is the SPM solved using finite-difference methods. We also investigate the influence of labeled data on model accuracy and generalization performance. All training experiments were conducted on an NVIDIA RTX 4500 Ada Generation GPU.

*4.1 PI-SPM under constant current condition*

First, the three models were evaluated under constant-current conditions. The training currents were

$$I_{\text{train}} = [1, 2, 4, 6, 8, 10] \text{ A},$$

and the test currents were

$$I_{\text{test}} = [1.5, 3, 5, 7, 9] \text{ A}.$$

To ensure a relatively fair comparison, the number of neurons in each model was adjusted such that all three network architectures contained approximately the same number of trainable parameters (about 6,650). In addition, all training hyperparameters and sampling strategies were kept identical. The models were trained using the Adam optimizer with an initial learning rate of $1 \times 10^{-4}$, followed by a 90% exponential decay every 1,500 epochs. The number of samples used for the two boundary conditions and the PDE residual were 200, 500, and 2,000, respectively. The total number of training epochs was 20,000.

The performance of the three models on the test data is shown in Fig. 3. The accuracy of the split and merged architecture shows no significant difference. The DeepONet-based model exhibits larger errors and requires longer training time. However, this does not imply that the DeepONet architecture is inferior to the other two. Because the three architectures differ substantially, their optimal hyperparameters may not be identical. Although DeepONet shows the worst relative performance among the three, its predicted concentrations and terminal voltages still agree well with the finite-difference-based reference solutions, as illustrated in Fig. 4.

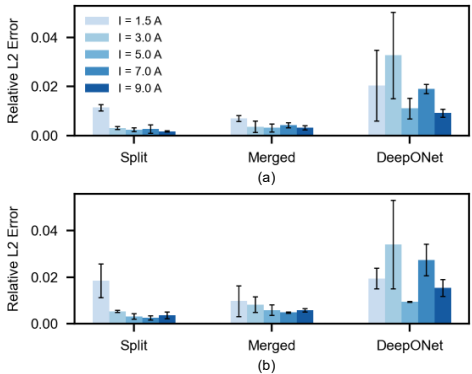

Figure 3. Relative $L_2$ error of the three models under the test conditions. (a) Positive-electrode concentration error. (b) Negative-electrode concentration error.

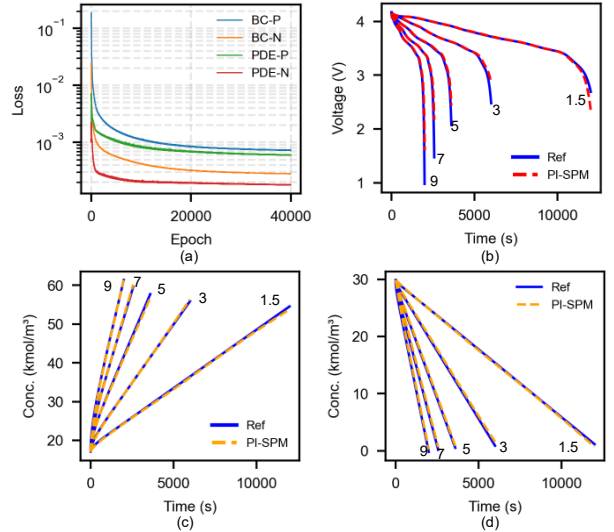

Figure 4. Performance of the PI-SPM (DeepONet) model under the test conditions. (a) Boundary-condition (BC) loss and PDE loss. (b) Terminal voltage. (c) Positive-electrode surface concentration. (d) Negative-electrode surface concentration.

Furthermore, we evaluated the model performance under current conditions beyond the upper boundary of the training data (10 A) to test the model's extrapolation capability, where the test currents of 11 A and 13 A were selected. The results indicate that the extrapolation performance of the split and merged architectures shows no significant difference. Both models produce accurate predictions at 11 A but exhibit noticeable degradation at 13 A. These observations suggest that the models possess a certain degree of extrapolation capability, but only within a limited range.

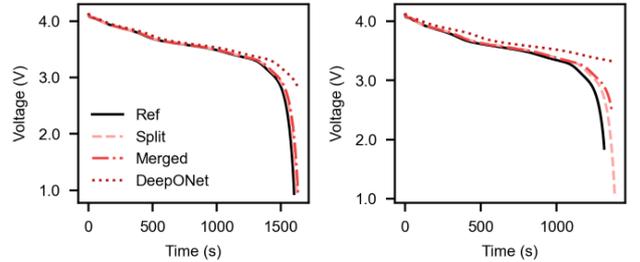

Figure 5. Model performance under extrapolation conditions (left: 11 A, right: 13 A).

*4.2 PI-SPM under dynamic current condition*

In real applications, discharge currents can vary arbitrarily, and it is impossible to cover all possible current profiles in the training data. As a result, learning the relationship between highly dynamic current inputs and the resulting concentration distributions becomes particularly challenging for neural-network-based surrogate models. To simplify the task, we first consider a low-frequency dynamic current profile generated using Gaussian Random Fields (GRF), with the current magnitude ranging from 1 to 10 A.

Fig. 6 illustrates the training results of three architectures under only one dynamic current in the train dataset. Among the three models, only the DeepONet-based PI-SPM performs well under the GRF current condition. The split and merged architectures struggle to learn the PDE solution under dynamic boundary conditions without access to labeled data. As shown in Figs. 6(c) and 6(d), although both the boundary-condition loss and PDE loss decrease during training, the prediction errors for the positive and negative electrode concentrations continue to increase. Even after increasing the number of samples, the number of neurons, the weighting of the boundary-condition loss, and even after applying curriculum regularization (Krishnapriyan et al., 2021). These two architectures consistently exhibit similar failure patterns.

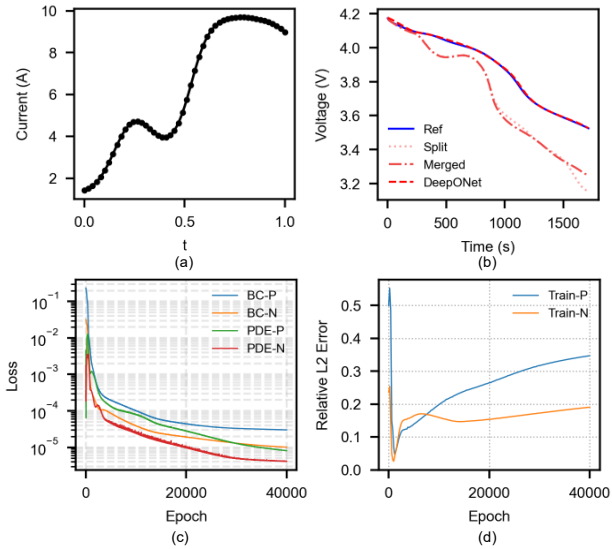

Figure 6. Training results of the three models under GRF current conditions. (a) Sample GRF current profile. (b) Voltage curves. (c) Loss curves of the merged PI-SPM during training. (d) Relative $L_2$ error of the merged PI-SPM concentrations during training.

The performance of the DeepONet-based PI-SPM on the test dataset is shown in Fig. 7. There are 100 current samples in the training dataset and 50 current samples in the test dataset. As illustrated in Fig. 8, its generalization ability strongly depends on the number of current samples used during training. Achieving satisfactory generalization on the test data requires incorporating hundreds of GRF current samples into the training dataset. However, this substantially increases the computational cost. For example, training with 100 current samples for 20,000 epochs takes approximately 15 hours.

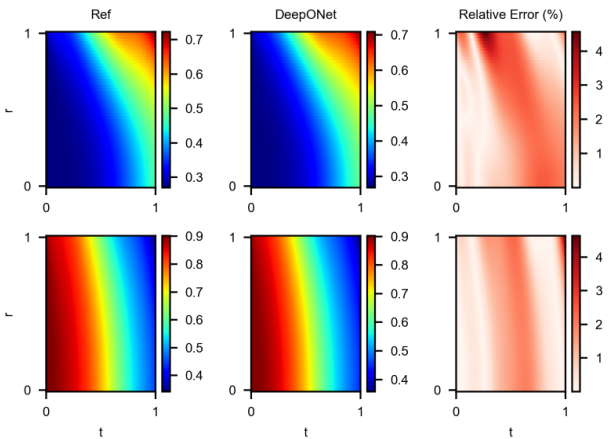

Figure 7. Performance of the DeepONet-based PI-SPM under GRF test current conditions. (Top: positive-electrode concentration; bottom: negative-electrode concentration).

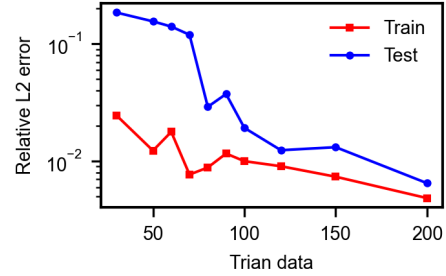

Figure 8. Error trend with different numbers of training current samples.

The GRF current profile varies only at very low frequencies, making the operator relatively easy to learn but not representative of real applications. In practice, the current in energy-storage systems and electric vehicles is highly dynamic and contains frequent abrupt changes. To better capture such behavior, we introduce the Random Step Current (RSC) profile. The RSC is a piecewise-constant waveform with randomly sampled amplitudes and abrupt transitions, resulting in a highly dynamic, non-periodic current profile.

Fig. 9(a) shows an example of an RSC profile, and Fig. 9(b) compares the two electrodes' surface concentrations predicted by the DeepONet-based PI-SPM with the reference solution. The results indicate that DeepONet can follow the general trend but fails to capture the high-frequency components. This limitation is closely related to the spectral bias of fully connected neural networks, a well-known pathology that prevents PINN-type models from accurately learning high-frequency functions (Wang et al., 2021b). Wang et al. (2021b) analyzed this issue using Neural Tangent Kernel theory and proposed two multi-scale architectures: the multi-scale Fourier feature architecture and the spatio-temporal multi-scale Fourier feature architecture.

In the spatio-temporal multi-scale Fourier feature method, different frequency bands can be selected according to the characteristics of the target problem. In this work, we use $\sigma_r = 1$ and $\sigma_t = 30$ to help the network better learn high-frequency dynamics in the time domain. As shown in Fig. 9(c), after applying the Fourier feature mapping, the predicted surface concentrations exhibit significantly improved agreement with the dynamic features of the reference solution.

We further compare the computation time of the traditional finite-difference solver (with 50 discretization points in the radial domain) and the Fourier-enhanced DeepONet surrogate model under the RSC current profile. For 50 repeated simulations on a CPU, the average computation times are 551 ms for the finite-difference method and 56 ms for the surrogate model. This corresponds to approximately a 10× speedup. Such an acceleration becomes particularly advantageous in scenarios requiring repeated simulations, including sensitivity analysis, parameter identification, and real-time control, where rapid evaluation of battery models is essential.

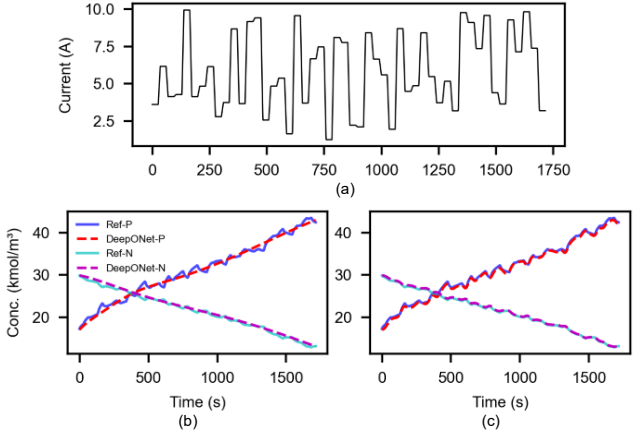
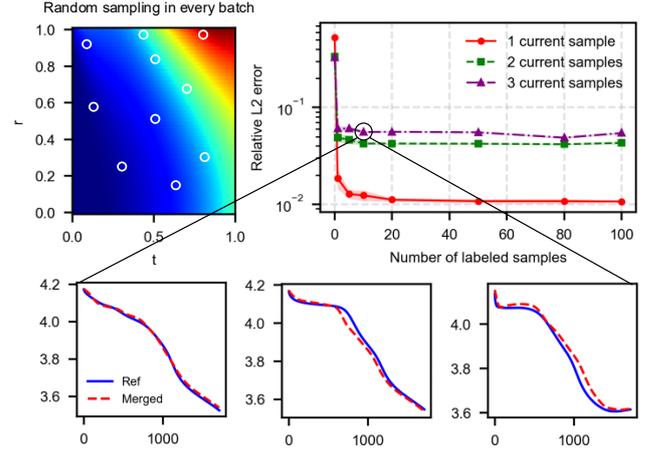

Figure 9. Model performance under high-frequency current conditions. (a) A sample of the RSC profile. (b) Surface concentrations without Fourier feature mapping. (c) Surface concentrations with Fourier feature mapping.

*4.3 Model evaluation on labeled data*

In the previous section, when trained without labeled data, the split and merged PI-SPM models failed to learn the correct mapping between the GRF current input and the concentration distribution. The central idea of physics-informed modeling is to combine data information with physical constraints to achieve both accurate and generalizable predictions. However, studies in the existing literature often rely exclusively on either the physics-based loss or the data-based loss, and few have investigated how incorporating labeled data can improve model performance. In this section, we examine the influence of labeled data on training performance, focusing on the merged PI-SPM architecture.

During training, a specified number of labeled samples are randomly selected from the reference dataset in each batch and incorporated into the loss function as the data loss term. We evaluate the merged PI-SPM under different numbers of training current profiles and different numbers of labeled samples. The results are shown in Fig. 10.

The observations reveal several important trends. First, even a single labeled sample is sufficient to guide the neural network toward the correct solution. Increasing the number of labeled samples beyond approximately ten does not yield further accuracy improvements, indicating diminishing returns. Second, the merged PI-SPM achieves very low error when trained on a single current profile. However, when trained on two current profiles, its error increases substantially, and the degradation is even more pronounced when trained on three profiles. Compared with the DeepONet-based PI-SPM in Fig. 8, this highlights a fundamental limitation of conventional PINN-based architectures: while they can achieve high accuracy for a single PDE with fixed boundary conditions, they struggle to learn a family of PDEs with varying boundary conditions—demonstrating weaker operator-learning capability than DeepONet.

Figure 10. Training results of the merged PI-SPM with different numbers of labeled samples.

## 5. CONCLUSIONS

This work systematically evaluates three physics-informed surrogate architectures for the SPM, namely split PI-SPM, merged PI-SPM, and DeepONet-based PI-SPM. Under constant-current conditions, all models achieve good accuracy and proper extrapolation capability. However, when dynamic current profiles are introduced, fundamental differences emerge. The conventional PINN-based architectures consistently fail to learn PDE solutions under dynamic boundary conditions without labeled data, whereas DeepONet-based PI-SPM remains effective but requires hundreds of training current samples to generalize properly. For highly dynamic inputs such as Random Step Currents, DeepONet exhibits spectral-bias limitations and struggles to capture high-frequency concentration dynamics. Incorporating spatio-temporal Fourier feature mappings significantly mitigates this issue, enabling accurate prediction of fast transients. To demonstrate the data efficiency, we also investigate how different amounts of labeled data influence model accuracy.

Overall, this study clarifies the operating regimes under which PINN-based SPM surrogates remain feasible and identifies Fourier-enhanced DeepONet as a more robust approach for realistic, highly dynamic battery-operation scenarios. These insights provide a practical foundation for developing more generalizable physics-informed surrogate models for battery-management applications.

Several important questions remain open for future research. The SPM is applicable only at low C-rates (≤1C), and constructing surrogate models for the P2D or even multi-physics frameworks with strong generalization remains highly challenging. Physics-encoded models (Faroughi et al., 2024; Huang et al., 2024) may offer promising directions in this regard. Furthermore, model parameters may vary as functions of temperature, stress, and aging. How to enable surrogate models to capture parameter variability is another important challenge. Although some studies (Brendel et al., 2025; Hassanaly et al., 2024b; Panahi et al., 2025) have included selected parameters as neural-network inputs, physics-based battery models typically contain more than twenty parameters, making comprehensive parameter conditioning a difficult yet crucial research topic.

## ACKNOWLEDGEMENTS

This work was supported in part by the Villum Foundation for Smart Battery project (No. 222860). Yunhong Che acknowledged the Novo Nordisk Foundation (No. NNF24OC0088261).

## DECLARATION OF GENERATIVE AI AND AI-ASSISTED TECHNOLOGIES IN THE WRITING PROCESS

During the preparation of this work, the authors used ChatGPT to assist in language polishing and clarity improvement. All technical content, analyses, and conclusions were developed solely by the authors, who take full responsibility for the final manuscript.

## Appendix A. FIRST APPENDIX

Table A1 SPM parameters

| Parameter | Unit | Description | Positive electrode | Negative electrode |
|---|---|---|---|---|
| $L$ | m | Electrode thickness | 75.6e-6 | 85.2e-6 |
| $R_s$ | m | Particle radius | 5.22e-6 | 5.86e-6 |
| $\varepsilon_s$ | | Active material volume fraction | 0.665 | 0.75 |
| $D_s$ | m$^2$s$^{-1}$ | Diffusion coefficient | 4.0e-15 | 2.0e-14 |
| $k$ | A m$^{2.5}$mol$^{-1.5}$ | Reaction rate constant | 1.0e-5 | 1.0e-5 |
| $c_{s,max}$ | mol m$^{-3}$ | Maximum solid-phase concentration | 63104 | 33133 |
| $c_{e,0}$ | mol m$^{-3}$ | Initial electrolyte concentration | 1000 | |
| $A$ | m$^2$ | Electrode coating area | 0.1027 | |
| $T$ | K | Temperature | 298.15 | |
| $F$ | C mol$^{-1}$ | Faraday's constant | 96485 | |
| $R$ | J mol$^{-1}$K$^{-1}$ | Universal gas constant | 8.3145 | |